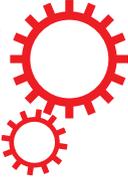

# Signature of a continuous quantum phase transition in non-equilibrium energy absorption: Footprints of criticality on higher excited states

Sirshendu Bhattacharyya[1], Subinay Dasgupta[2] & Arnab Das[3]

Understanding phase transitions in quantum matters constitutes a significant part of present day condensed matter physics. Quantum phase transitions concern ground state properties of many-body systems, and hence their signatures are expected to be pronounced in low-energy states. Here we report signature of a quantum critical point manifested in strongly out-of-equilibrium states with finite energy density with respect to the ground state and extensive (subsystem) entanglement entropy, generated by an external pulse. These non-equilibrium states are evidently completely disordered (e.g., paramagnetic in case of a magnetic ordering transition). The pulse is applied by switching a coupling of the Hamiltonian from an initial value ($\lambda_I$) to a final value ($\lambda_F$) for sufficiently long time and back again. The signature appears as non-analyticities (kinks) in the energy absorbed by the system from the pulse as a function of $\lambda_F$ at critical-points (i.e., at values of $\lambda_F$ corresponding to static critical-points of the system). As one excites higher and higher eigenstates of the final Hamiltonian $H(\lambda_F)$ by increasing the pulse height ($|\lambda_F - \lambda_I|$), the non-analyticity grows stronger monotonically with it. This implies adding contributions from higher eigenstates help magnifying the non-analyticity, indicating strong imprint of the critical-point on them. Our findings are grounded on exact analytical results derived for Ising and XY chains in transverse field.

A continuous quantum phase transition (henceforth QPT) entails non-analytic behaviour of the ground-state properties of a system at the quantum critical point[1–4]. QPT is observed ubiquitously in diverse areas of condensed matter physics and has extensively been studied through last few decades[3]. Interest in different aspects of it, with many important open issues is nevertheless growing rapidly with time.

One important and long-standing question concerns the existence of the fingerprints of a QPT on higher excited states (i.e., states with finite energy density with respect to the ground state). It is known that QPT has definitive signatures in the ground-state (and low-lying states close to it), as reflected in certain ground-state properties like the derivative of ground-state entanglement (concurrence)[5], the ground-state fidelity susceptibility (see, e.g.,[6–11]) or scaling of ground-state entanglement entropy at the critical point(see, e.g.,[12–14]). For a physical (local) Hamiltonian, it is not unexpected for the states moderately close to the ground state (with vanishing energy density) to retain such direct mark of the singularity on them. But as one goes up in energy scale (and hence down in length-scale) the non-universal microscopic details of the system tend to dominate most of the physical properties of the system, and the signature of the ground-state singularity gets obscured. This however does not rule out existence of such a signature lying hidden in those excited states in some form.

[1]R.R.R. Mahavidyalaya, Radhanagar, Hooghly 712406, India. [2]Department of Physics, University of Calcutta, 92 Acharya Prafulla Chandra Road, Kolkata 700009, India. [3]Theoretical Physics Department, Indian Association for the Cultivation of Science, 2A & 2B Raja S. C. Mullick Road, Kolkata - 700032, India. Correspondence and requests for materials should be addressed to A.D. (email: arnab.das.physics@gmail.com or tpad5@iacs.res.in)

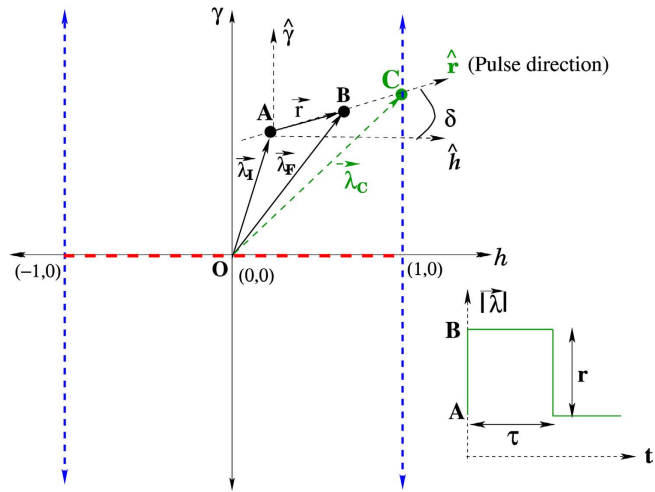

**Figure 1. Phase diagram of XY chain in transverse field at $T=0$ in the $h-\gamma$ plane.** The three critical lines (in dashes) are shown - the anisotropy transition in red (along $\gamma=0$, between $h=\pm 1$) and the Ising transition lines in blue. Application of an arbitrary pulse in the $h-\gamma$ plane is illustrated.

From the point of view of equilibrium physics, a faint positive indication follows from the existence of the so-called quantum critical region in the finite temperature phase diagram of a quantum system, where the behaviors of the system is strongly affected by the existence of an underlying quantum critical point even at $T\neq 0$[3]. Similar indication is obtained from non-equilibrium response mechanism of a system driven at a finite rate across a quantum critical point, largely known as the Kibble-Zurek mechanism[15,16], which predicts universal scaling behaviour of defect density with respect to the drive rate. It has been observed that similar to the classical case[17,18], the said universality holds even for rapid drives that produces finite density of non-adiabatic excitation also in the quantum case[19–24].

However, a clear signature of quantum criticality on high-lying states has remained elusive so far. Here we bridge this gap by demonstrating a case where the fingerprints of the QPT on the higher-excited states play an *essential* role in the occurrence of a non-analyticity in the non-equilibrium response of the system. The signature of the critical point is observed in the energy absorbed by the system from a pulse of large duration. Stronger signature is observed as contribution from higher excited states are incurred by applying pulse of larger amplitude. Though a non-analyticity does not necessarily imply a phase transition (see, e.g.[25], where non-analytic behaviour of correlation length is observed at a point with no phase transition), in our case the signature always and exclusively occurs at the critical points.

### Results

Consider a quantum many-body Hamiltonian $H(\vec{\lambda})$ with a set of $d$ independent coupling parameters defining a $d$-dimensional parameter space, where a position vector $\vec{\lambda}$ represents a set of values for the couplings. We start with the initial state $|\psi(0)\rangle$ to be the ground-state $|\varepsilon_0^I\rangle$ of $H(\vec{\lambda}_I)$ with eigenvalue $\varepsilon_0^I$. At $t=0$ we apply a pulse of strength $r$ in a given direction $\hat{r}$ by switching the value of $\vec{\lambda}$ from $\vec{\lambda}_I$ to $\vec{\lambda}_F = \vec{\lambda}_I + r\hat{r}$, and again at $t=\tau$, switch it back to $\vec{\lambda}_I$ (see Fig. 1). The initial energy of the system at $t=0$ is the ground-state energy $\varepsilon_0^I$. The final energy of the system (after the pulse is withdrawn) is $E^F = \langle\psi(\tau)|H(\vec{\lambda}_I)|\psi(\tau)\rangle$. The absorbed energy is hence

$$E_{\text{abs}} \equiv E^F - E^I = \langle\psi(\tau)|H(\vec{\lambda}_I)|\psi(\tau)\rangle - \varepsilon_0^I. \tag{1}$$

We show (our main result), that *if and only if* for a particular value of $r$ ($r=r_c$), $\vec{\lambda}_F = \vec{\lambda}_C$ corresponds to quantum phase transition point, then the first derivative of $E_{\text{abs}}$ with respect to $r$ is discontinuous at $r=r_c$ (a general mathematical argument is given under the heading "General Outline of our Calculation"). Henceforth we drop the arrows and hats when unambiguous. Note that it is actually sufficient to study the final energy of the system instead of $E_{\text{abs}}$ but energy transfer is usually easier to measure than absolute energy.

For the sake of illustration we consider the XY-chain in a transverse field:

$$H = -\frac{J}{2}(1+\gamma)\sum_{j=1}^{N}\sigma_j^x\sigma_{j+1}^x - \frac{J}{2}(1-\gamma)\sum_{j=1}^{N}\sigma_j^y\sigma_{j+1}^y - h\sum_{j=1}^{N}\sigma_j^z \qquad (2)$$

where $\sigma^{x,y,z}$ represents the three components of a Pauli spin, $\gamma$ and $h$ denote the anisotropy and the transverse field respectively (we take $\hbar = J = 1$). The $T = 0$ phase diagram of the system has three critical lines as shown in (Fig. 1). Pulses of varying height $r$ along a chosen direction $\hat{r}$ in the $h$-$\gamma$ plane are applied by changing parameters from $\vec{\lambda}_I \equiv (h_I, \gamma_I)$ to $\vec{\lambda}_F \equiv \vec{\lambda}_I + r\hat{r} \equiv (h_F, \gamma_F)$, where $r = \sqrt{(h_F - h_I)^2 + (\gamma_F - \gamma_I)^2}$, $h_F - h_I = r \cos\delta$, $\gamma_F - \gamma_I = r \sin\delta$, $\delta$ being the angle between $\hat{r}$ and the positive $h$-axis (see Fig. 1).

The general expression for $E_{abs}$ can be obtained as follows. Hamiltonian (2) can be written as direct product of $N/2$ mutually commuting operators $\mathcal{H}_k$, each acting on a separate 2-D Hilbert space spanned by occupation number states $|0_k, 0_{-k}\rangle$ and $|1_k, 1_{-k}\rangle$, in the momentum ($k$)-space (see, e.g.,[3,4]). $H_k$ has eigenvalues $\pm\epsilon_k(h,\gamma) = \pm 2\sqrt{(\cos k + h)^2 + \gamma^2 \sin^2 k}$ corresponding respectively to eigenvectors $|\epsilon_k\rangle_+ = i \sin\theta|1_k, 1_{-k}\rangle + \cos\theta|0_k, 0_{-k}\rangle$ and $|\epsilon_k\rangle_- = i \cos\theta|1_k, 1_{-k}\rangle - \sin\theta|0_k, 0_{-k}\rangle$, where $\tan\theta = -(\gamma \sin k)/[\cos k + h + \sqrt{(\cos k + h)^2 + \gamma^2 \sin^2 k}]$. We start at $t = 0$ with the ground-state of $H(h_I, \gamma_I)$, given by $|\psi(t=0)\rangle = \prod_k |\varepsilon_k^I\rangle_-$ (where super/subscript $I$ and $F$ indicate values corresponding to $(h_I, \gamma_I)$ and $(h_F, \gamma_F)$ respectively. One can express: $|\epsilon_k^I\rangle_- = \cos\phi|\epsilon_k^F\rangle_- + \sin\phi|\epsilon_k^F\rangle_+$, where $\phi = \theta_F - \theta_I$. Thus, for $0 < t < \tau$, the wave function is $|\psi(t)\rangle = \prod_k [\cos(\phi) e^{i\epsilon_k^F t}|\epsilon_k^F\rangle_- + \sin(\phi) e^{-i\epsilon_k^F t}|\epsilon_k^F\rangle_+]$. Similarly for $t > \tau$, $|\psi(t)\rangle = \prod_k [(\cos\mu + i \sin\mu \cos 2\phi)|\epsilon_k^I\rangle_- + i \sin\mu \sin 2\phi|\epsilon_k^I\rangle_+]$, where $\mu = (\epsilon_k^F)\tau$ (see, e.g.[26]). The absorbed energy as defined in Eq. (1) is thus $E_{abs} = \frac{1}{\pi}\int_0^\pi \epsilon_k^I \sin^2 2\phi \sin^2 \mu \, dk$ in the $N \to \infty$ limit. Now we consider the so called dephasing limit $\tau \gg |J|, |h|$, in which all phases that include $\tau$ can be considered rapidly oscillating and are dropped out from the calculation of any physical quantity following Riemann-Lebesgue Lemma (see, e.g.,[27]). In this limit $\sin^2\mu = (1 - \cos 2\mu)/2$ is hence replaced by $\frac{1}{2}$. Using this, and expressions for $\epsilon_k$ and $\phi(\theta)$, we get

$$E_{abs} = \frac{1}{\pi}\int_0^\pi \frac{\widetilde{W}(h_I, \gamma_I, h_F, \gamma_F)}{\sqrt{\widetilde{Q}(h_I, \gamma_I)}\,\widetilde{Q}(h_F, \gamma_F)} dk, \qquad (3)$$

where $\widetilde{W} = [\gamma_I(h_F + \cos k) - \gamma_F(h_I + \cos k)]^2 \sin^2 k$, $\widetilde{Q}(h,\gamma) = (\cos k + h)^2 + \gamma^2 \sin^2 k$ and the square root refers to the principal branch. Our quantity of interest is $\left[\frac{\partial E_{abs}}{\partial r}\right]_{\vec{\lambda}_I, \delta}$ and our main result is that the derivative changes discontinuously as a function of $r$ at $r = r_c$. The analytic expression for the measure of discontinuity (size of the jump in the derivative) is given by

$$\Delta \equiv \left(\frac{\partial E_{abs}}{\partial r}\right)_- - \left(\frac{\partial E_{abs}}{\partial r}\right)_+,$$

(where the subscripts $\mp$ refer respectively to the values $r = r_c \mp \eta$, with $\eta \to 0^+$) which is the quantitative measure of the non-analyticity.

**General Outline of our Calculation.** The occurrence of the non-analyticity can be qualitatively seen from the general outline of our calculation given in the following. We switch to the complex plane by substituting $z = e^{ik}$ in the above expression and convert the integral in Eq. (3) to one over a circular contour $\mathcal{C}$ with unit radius centering the origin in the $z$-plane:

$$E_{abs} = \frac{i}{4\pi}\mathcal{A}\oint_\mathcal{C} \frac{W(z, h_I, \gamma_I, h_F, \gamma_F)}{\sqrt{Q(z, h_I, \gamma_I)}\,z^2 Q(z, h_F, \gamma_F)} dz, \qquad (4)$$

where

$$\mathcal{A} = \frac{1}{(1-\gamma_F^2)\sqrt{1-\gamma_I^2}},$$

$$W = (z^2 - 1)^2[\gamma_I(z^2 + 2h_F z + 1) - \gamma_F(z^2 + 2h_I z + 1)]^2,$$

$$Q(z, h, \gamma) = (z - z_1)(z - z_2)\left(z - \frac{1}{z_1}\right)\left(z - \frac{1}{z_2}\right)$$

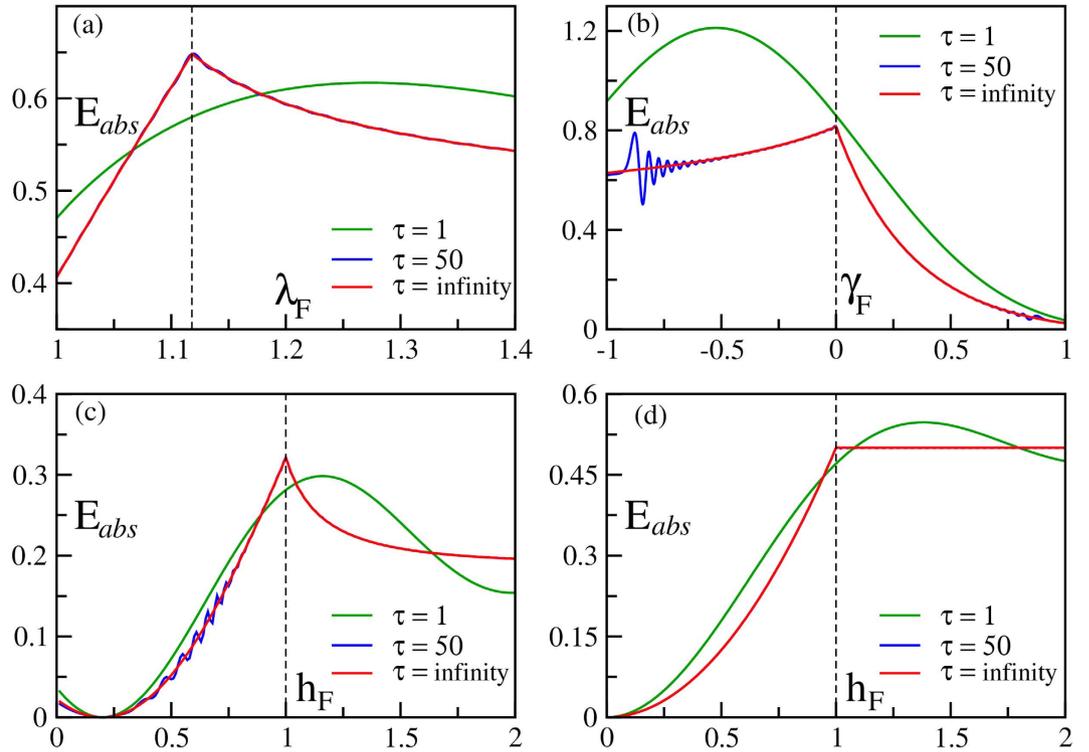

**Figure 2. Kink signatures for infinite and finite $\tau$.** Plot of $E_{abs}$ for different types of pulses are shown. Clear signatures (kinks) are visible at critical points marked by vertical dotted lines. (**a**) $E_{abs}$ vs $\lambda_F$ ($|\vec{\lambda}_F|$) plot for a general type of pulse ($\lambda_I \to \lambda_F \to \lambda_I$) starting from $h_I = 0$, $\gamma_I = 1$ in the direction given by $\tan\delta = -0.5$. The critical point occurs at $\lambda_F = 1.118$. (**b**) $E_{abs}$ vs $\gamma_F$ plot for an anisotropic pulse starting from $\gamma_I = 1.5$ with fixed $h = 0.2$. (**c**) $E_{abs}$ vs $h_F$ plot for a field-pulse starting from $h_I = 0.2$ with fixed anisotropy, $\gamma = 0.5$. (**d**) $E_{abs}$ vs $h_F$ plot for a transverse Ising model with a transverse field-pulse $0 \to h_F \to 0$. For $\tau = 50$, the plots are almost indistinguishable from the infinite $\tau$ result, particularly at and around the critical point.

with

$$z_{1,2}(h, \gamma) = \frac{1}{1-\gamma}\left[-h \pm \sqrt{h^2 + \gamma^2 - 1}\right]. \tag{5}$$

Note that $W$ is analytic within $\mathcal{C}$, $\sqrt{Q}$ brings in branch-points into the problem which can be avoided by drawing proper branch-cuts and indenting the contour so that the only singularities enclosed within $\mathcal{C}$ are the poles, coming from the zeros of $z^2 Q$. The non-analytic behaviour of $E_{abs}$ with respect to $r$ occurs because the pole structure within $\mathcal{C}$ are different for $r > r_c$ and $r < r_c$ – poles move in/out of the contour as $r$ is varied across $r_c$. Thus the functional dependence of $E_{abs}$ on $r$ (which depends on the residues of the poles within $\mathcal{C}$) are in general different on two sides of $r_c$, giving rise to the non-analyticity at (and only at) the critical point. In this context it is interesting to note that *no such non-analyticity is observed in $E_{abs}$ (see Suppl. Mat.) for a single quench $H(\lambda_I) \to H(\lambda_F)$ as a function of $\lambda_F$* in an XY chain (see e.g.[28–30]). In the following we outline our approach for calculating $\Delta$ exactly analytically for pulses in arbitrary directions in the parameter plane.

**Crossing the Ising ($h=1$) Critical Line.** Here we consider pulses from a general initial point ($h_I, \gamma_I$) towards a direction $\delta$ such that the Ising transition line is crossed (e.g., in Fig. 2c). Evaluating the integral for $E_{abs}$ (Eq. 4) following the general outline given above (see the Method section for details) we get, for $h_F = 1$, $\gamma_F \neq 0$,

$$\Delta = \frac{(1-h_I)}{|\gamma_I + m(1-h_I)|\sqrt{1+m^2}}, \tag{6}$$

where $m = \tan(\delta)$. $\Delta = 0$ for any other point on the pulse-line that is not a QPT point. This is of course under the assumption that we are considering the nearest point of intersection between the pulse-line and the critical line from the initial point. If the line intersects more than one critical lines, $E_{abs}$ shows discontinuity ($\Delta \neq 0$) at each of those (see Fig. 3). A further simplification occurs if we consider the Ising case, i.e., $\gamma = 1$, under the $h$-pulse $0 \to h_F \to 0$, where we can calculate the entire analytical expression: $E_{abs} = \left(\frac{h_F^2}{2}\right)$ for $h_F \leq 1$, and $\left(\frac{1}{2}\right)$ for $h_F \geq 1$ (Fig. 2d).

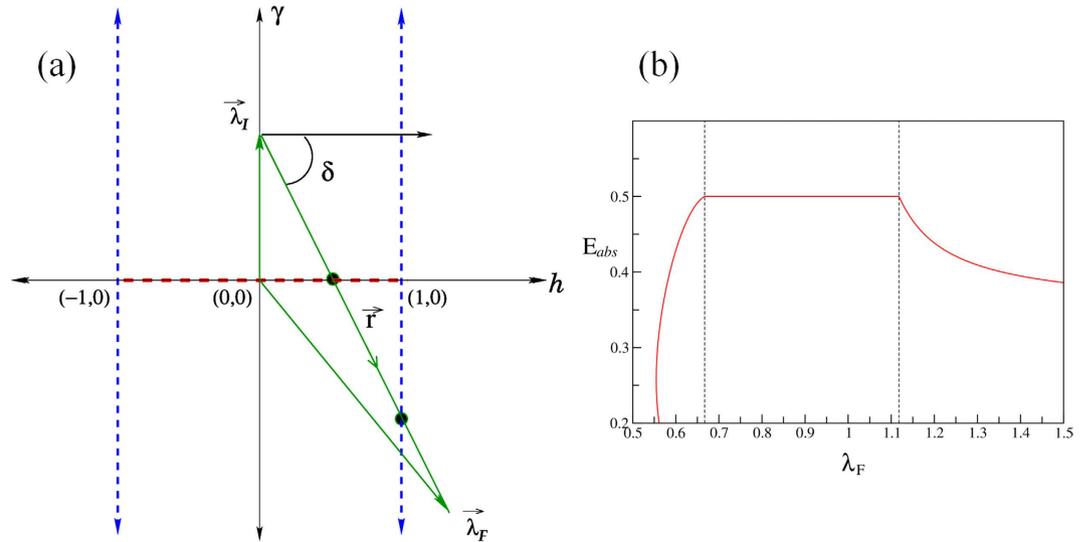

**Figure 3.** (a) Pulse line intersecting two critical lines. This happens for $-\frac{3\pi}{4} < \delta < -\frac{\pi}{4}$. For this figure, $\delta = -\tan^{-1}(1.5)$. (b) Corresponding $E_{abs}$ vs $\lambda_F$. The present case corresponds to initial point $\lambda_I = (h_I = 0, \gamma_I = 1)$, the pulse direction $\tan\delta = -1.5$, and pulse duration $\tau \to \infty$. The pulse-line crosses the two critical lines at $\lambda_F = 0.667$ and $\lambda_F = 1.118$ which are the anisotropy critical line and the Ising critical line respectively. The non-analyticity is observed in both the cases.

**Crossing the Anisotropic ($\gamma = 0$) Critical Line.** In this case (as in Fig. 2b), when we cross $\gamma_F = 0$ critical line in the region $-1 < h < 1$, the size of discontinuity is given by

$$\Delta = \frac{2\gamma_I[1 + \gamma_I(\gamma_I - 2mh_I) - m^2(1 - h_I^2)]}{\sqrt{1 - \gamma_I(\gamma_I - 2mh_I) + m^2(1 - h_I^2)}\sqrt{1 + m^2}} \quad (7)$$

Note that for $|h| > 1$, the $\gamma = 0$ line is *not* critical since there is no restructuring of the locations of poles (see Methods section).

**The Multi-Critical Points (MCP)** are characterized by a discontinuity in $\Delta$ as a function of $\delta$. Suppose we keep $(h_I, \gamma_I)$ fixed and vary the pulse direction $\delta$, then $\Delta(\delta)$ is *discontinuous* at the values of $\delta$ for which the pulse-direction passes through the MCPs at $(h = \pm 1, \gamma = 0)$. For example, choosing the initial point to be $(h_I, \gamma_I) = (0, 1)$, we get the following values of $\Delta$ depending on $\delta$ within the infinitesimal neighborhood of the *MCP*.

$$h_F = 1 - \eta, \quad \gamma_F = 0; \quad \delta = -(\pi/4 + 0^+) \Rightarrow \lim_{\eta \to 0^+} \Delta \to 0$$

$$h_F = 1, \quad \gamma_F = 0; \quad \delta = -\pi/4 \quad \Rightarrow \Delta = -\frac{1}{\sqrt{2}} (MCP)$$

$$h_F = 1, \quad \gamma_F = 0 + \eta; \quad \delta = -(\pi/4 - 0^+) \Rightarrow \lim_{\eta \to 0^+} \Delta \to \infty$$

**Crossing More than one critical lines.** In some cases, as one increases $r$ $(= |\vec{\lambda}_F - \vec{\lambda}_I|)$, one crosses both the anisotropic critical line and the Ising critical line (Fig. 3a) so that the $E_{abs} - \lambda_F$ curve has two kinks (Fig. 3b) due to (independent) changes in the pole structure twice corresponding to the crossing of two critical points (see Methods Sec. for details).

**Signature on the Higher Excited States.** Here we show that the non-analyticity in $E_{abs}$ at $\lambda_F = \lambda_C$ essentially involves contributions from higher excited states of $H(\lambda_I)$ and $H(\lambda_C)$. Let $\{|\varepsilon_n^I\rangle\}$ and $\{|\varepsilon_n^F\rangle\}$ denote the complete set of ortho-normalized eigenstates of $H(\lambda_I)$ and $H(\gamma_F)$, and $\chi_{mn} = \langle \varepsilon_m^F | \varepsilon_n^I \rangle$. It can be shown that $E_{abs} = |\chi_{00}|^4 \varepsilon_0^I + \sum_{n \neq 0, n'}(|\chi_{n0}|^2 |\chi_{nn'}|^2 \varepsilon_n^I) - \varepsilon_0^I$. The overlap $|\chi_{00}| = |\langle \varepsilon_0^F | \varepsilon_0^I \rangle|$ vanishes exponentially with $N$ (see, e.g., Ref. 31), and $\varepsilon_0^I$ is independent of $N$ Hence the "ground-state-only" con-

tribution $|\chi_{00}|^4 \varepsilon_0^I$ vanishes as $N \to \infty$ (note that $|\chi_{00}|$ is nothing but the fidelity between $|\varepsilon_0^I\rangle$ and $|\varepsilon_0^F\rangle$.) Hence all the terms contributing to $E_{abs}$ (and hence its non-analyticity) involves higher excited states.

The question that naturally arises is how high can we go up in the energy and still find eigenstates (of the critical Hamiltonian $H(\lambda_C)$) contributing significantly to the signature? One quantitative way of probing this is to monitor the magnitude of the discontinuity $\Delta$ as the function of the pulse height $r$, as one increases it by moving $\vec{\lambda}_I$ away from the critical point deeper into a phase. An increase in $r$ will result in excitation of higher energy eigenstates of $H(\lambda_C)$, and hence more weight of such states in $E_{abs}$. Hence, if the signature weakens on the eigenstates of $H_c$ as one goes up in energy, then increase in $r$ is expected to decrease $\Delta$. But here $\Delta$ increases monotonically with such increase in $r$ without bound. For example, for crossing the Ising line (Eq. 6), we get $\Delta \propto |1 - h_I|$, and for crossing the anisotropy line (Eq. 7), $\Delta \propto |\gamma_I|$. This indicates that eigenstates with higher and higher energy (even those with non-vanishing energy density) excited by abitrary increase in $r$, actually carry the imprint of the critical point, which is filtered out and added up to enhance $\Delta$.

Finally, the state $|\psi(\tau)\rangle$ itself, from which we read our signature off, has extensive subsystem entanglement entropy (proportional to the linear dimension of the sub-system)[32,33]. Our signature is thus visible on a non-equilibrium state not only with extensive energy density, but also with extensive entanglement entropy - a state truly far from the ground state manifold.

**Signature in Transverse Spin Polarization.** $E_{abs}$ is not necessarily the only quantity from which the non-analytic signature of QPT can be read off. Here we illustrate this by discussing the special case of $0 \to h_F \to 0$ pulse, keeping $\gamma$ constant. The quantity we concentrate on is the total transverse spin polarization $M_z(t) \equiv \sum_j \sigma_j^z$. Though $[M_z(t), H(t)] \neq 0$, as $\tau \to \infty$ one gets $E_{abs} = h_F \langle M_z(\tau \to \infty)\rangle$, starting with a ground state of $H(h=0)$ as the initial state. Thus the non-analytic behaviour of $E_{abs}$ as a function of $h_F$ is directly reflected in the behaviour of $\lim \tau \to \infty M_z(\tau)$ vs $h_F$.

## Discussion

Though the models we have studied are integrable, we conjecture that $E_{abs}$ will continue to bear similar non-analytic signature of continuous QPT in generic systems, at least when $r_c$ is small. The intuition is based on the observation that $\Delta$ is a *continuous* function of $(h_I, \gamma_I)$ for a given critical point, i.e. if we move our initial point towards the transition point (thereby reduce $r_c$), the value of $\Delta$ reduces in a *continuous* manner. This means, even if our pulse height is small enough so that the ground state of $H(\lambda_I)$ is a superposition of only the low-lying eigenstates of $H(\lambda_c)$, still the discontinuity will persist. But appearance of non-analyticity at criticality due to excitation of only the low energy manifold of $H$ by small perturbations is expected to be universal. Hence existence of such non-analyticities in the tractable models suggests their existence in more generic systems. However, if indefinitely higher excited states will still continue to bear the signature in those cases is an open question.

The entire picture fits in if we assume that $E_{abs}(\vec{\lambda}_F)$, for a given $\vec{\lambda}_I$ and pulse direction is the property of the equilibrium "phase", i.e., it is a smooth function as long as the pulse-line between $\vec{\lambda}_I$ and $\vec{\lambda}_F$ lies within a single phase. If it crosses a phase boundary (transition line/point), it in general becomes a different smooth function. Hence $E_{abs}$ shows a non-analyticity at the boundary as a function of $\vec{\lambda}_F$.

The above picture allows us to consider $E_{abs}$ as a powerful substitute of order parameter for a quantum phase transition, with $\Delta$ serving as a sharp locator of the critical point. It is well known that identifying the correct order parameter for complex phases of a condensed matter may often prove to be a hard task. Locating a quantum critical point accurately can also be quite difficult. Our result indicates an alternative non-equilibrium route to the solution of these long-standing problems.

In conclusion, we demonstrate that quantum phase transition (QPT) can be detected from a truly non-equilibrium signature exhibited in the energy absorbed by the concerned system from an externally applied pulse. The non-analyticity essentially involves signature of the QPT on high-lying excited states. This opens up possibilities of devising protocols for detecting and confirming quantum critical points in systems where the location of the critical point and the nature of the phases across it are not precisely known.

Many open issues spring to mind. Do similar signatures exist for QPTs other than continuous ones (e.g., for 1st order) or those without local order parameters, like the topological QPTs? Since the signature involves higher excited states, does it persist at low but finite temperature? Finally, how high in energy can one find eigenstates bearing discernible signatures of the low-lying equilibrium quantum phase transition for a generic quantum critical system with *local* interactions? Of course, one can design a non-local Hamiltonian, say, by choosing independent orthogonal states as the eigenstates of the Hamiltonian, in which case higher excited states will, by construction, have no information whatsoever regarding the nature of the ground state. Hence the indication we obtain here delves into the deeper issues of correlation between the low-lying and higher eigenstates of a local Hamiltonian. These general open questions are likely to trigger further research in the field.

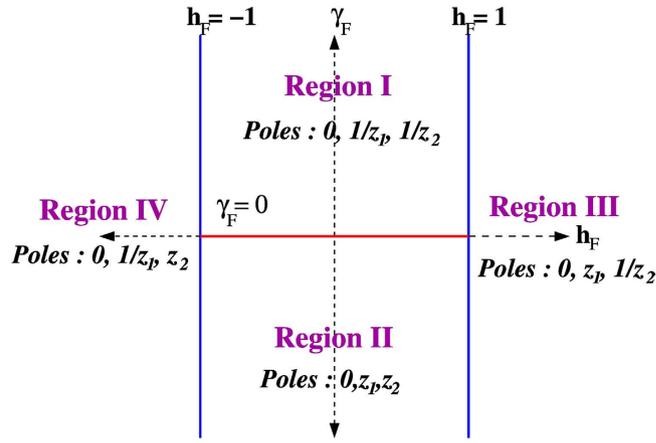

**Figure 4. Pole structure in the $h_F - \gamma_F$ plane.** The poles $z = 0, \frac{1}{z_1^F}, \frac{1}{z_2^F}$ are inside the contour for $|h_F| < 1$ and $\gamma_F > 0$ (named Region I); $z = 0, z_1^F, z_2^F$ are inside for $|h_F| < 1$ and $\gamma_F < 0$ (Region II); $z = 0, z_1^F, \frac{1}{z_2^F}$ are inside when $h_F > 1$ for all values of (positive and negative) $\gamma_F$ (Region III); $z = 0, \frac{1}{z_1^F}, z_2^F$ are inside when $h_F < -1$ for all values of $\gamma_F$ (Region IV).

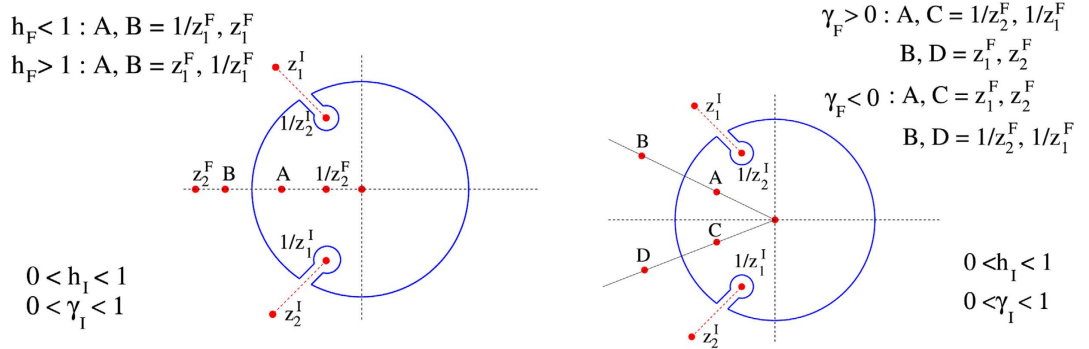

**Figure 5. Contours with poles and branch-point for integrating $E_{abs}$ (Eq. 4) under different circumstances:** The left contour represents the case of crossing the $h = 1$ line. $z^F$ corresponds to the arguments $h_F, \gamma_F$ and $z^I$ for $h_I, \gamma_I$. The right contour corresponds to the case of crossing the $\gamma = 0$. $z^F$ corresponds to the arguments $h_F, \gamma_F$ and $z^I$ for $h_I, \gamma_I$. In both the figures red dotted lines represent the branch-cuts.

## Methods

Here we lay down the details of evaluating $E_{abs}$ for different cases of crossing the critical lines. In the general expression Eq. (4) of $E_{abs}$, the evaluation of the integration of $E_{abs}$ involves the investigation of positions of poles and branch points in/outside the contour. There are four branch points: $z_{1,2}(h_I, \gamma_I)$, $\frac{1}{z_{1,2}(h_I, \gamma_I)}$ and five poles: 0 (of order 2), $z_{1,2}(h_F, \gamma_F)$, $\frac{1}{z_{1,2}(h_F, \gamma_F)}$ (of order 1). Branch points involving $(h_I, \gamma_I)$ and poles involving $(h_F, \gamma_F)$ will be denoted by $z^I$ and $z^F$ respectively. The positions of the poles change depending on the value of $h_F$ and $\gamma_F$. The critical lines $h_F = \pm 1$ and $\gamma_F = 0$ divide the $h_F - \gamma_F$ plane into four different regions where the poles residing inside the contour are different (elaborated in Fig. 4).

First we consider the case of crossing $h = 1$ critical line from positive $\gamma_I$ (Region I to Region III in Fig. 4). The pole $z = 0$ are always inside in all cases. Of the other four, $\frac{1}{z_2^F}$ is always inside and $z_2^F$ is always outside the contour. However, the pole $\frac{1}{z_1^F}$ is inside the unit circle for $h_F < 1$ and outside for $h_F > 1$. The branch points $\frac{1}{z_{1,2}^I}$ are inside the circle and the contour should be indented to avoid branch lines (Fig. 5, left contour). Since the integral over the indentations are continuous functions of $(h_F, \gamma_F)$, the non-analyticity in the behaviour of $E_{abs}$ arises from the restructuring of poles inside the unit circle and we have

$$E_{\text{abs}} = \text{A.P.} - \frac{\mathcal{A}}{2}Res. \left(\frac{1}{z_1^F}\right) \quad \text{for} \quad h_F < 1$$

$$E_{\text{abs}} = \text{A.P.} - \frac{\mathcal{A}}{2}Res. (z_1^F) \quad \text{for} \quad h_F > 1,$$

where A.P. stands for analytic part and Res. for residue of the integrand of $E_{\text{abs}}$. We then calculate the difference of slope of two sides of $h_F = 1$ to get Eq. (6).

Next we consider the case of crossing $\gamma = 0$ critical line. Here all the poles except $z = 0$ change their positions when $\gamma_F = 0$ is crossed [see Figs 4 and 5 (right contour)]. Thus we get

$$E_{\text{abs}} = \text{A.P.} - \frac{\mathcal{A}}{2}\left[Res. \left(\frac{1}{z_1^F}\right) + Res. \left(\frac{1}{z_2^F}\right)\right] \quad \text{for} \quad \gamma_F > 0$$

$$E_{\text{abs}} = \text{A.P.} - \frac{\mathcal{A}}{2}[Res. (z_1^F) + Res. (z_2^F)] \quad \text{for} \quad \gamma_F < 0,$$

This equation leads to the evaluation of the difference of slope before and after crossing $\gamma_F = 0$ as in Eq. (7). As mentioned above, the $\gamma = 0$ line is *not* critical for $|h| > 1$ because the poles $z_1^F, \frac{1}{z_2^F}$ remain inside and $\frac{1}{z_1^F}, z_2^F$ remain outside the unit circle for all values of $\gamma_F$. Hence, there is no restructuring of the location of poles, and no non-analyticity of absorbed energy in this case.

For pulses applied along a line that cross two critical lines/points, two independent kinks are observed at the two critical points. For example (Fig. 3b), when $\lambda_F$ crosses the $\gamma_F = 0$ line, the pole structure changes from $\left(0, \frac{1}{z_1^F}, \frac{1}{z_2^F}\right)$ inside to $(0, z_1^F, z_2^F)$ inside (see Fig. 4), and $\Delta$ is given by Eq. (7). Similarly when the pluse line crosses the $h_F = 1$ line, the pole $z_2^F$ goes outside and $\frac{1}{z_2^F}$ comes inside and $\Delta$ is given by Eq. (6).

### Acknowledgements
A. Das thanks R. Moessner, A. Sen, K. Sengupta, A.P. Young and W.H. Zurek for stimulating inputs, and the Max-Planck Institute (PKS) for hosting a visit during which a part of this work was done. The work of S. Dasgupta is supported by UGC-UPE (University of Calcutta).

### Author Contributions
All authors contributed to the theoretical analysis, interpretation of the results and the preparation of the manuscript. S.B. and S.D. carried out the major part of the computation. A.D. proposed the idea and steered the project.

### Additional Information
**Supplementary information** accompanies this paper at http://www.nature.com/srep

**Competing financial interests:** The authors declare no competing financial interests.

**How to cite this article**: Bhattacharyya, S. *et al.* Signature of a continuous quantum phase transition in non-equilibrium energy absorption: Footprints of criticality on higher excited states. *Sci. Rep.* **5**, 16490; doi: 10.1038/srep16490 (2015).